\documentstyle[aps,multicol,prl]{revtex}
\renewcommand{\narrowtext}{\begin{multicols}{2} \global\columnwidth20.5pc}
\renewcommand{\widetext}{\end{multicols} \global\columnwidth42.5pc}

\input{epsf.tex}

\def \attn #1 {{\sl $\bullet$ #1 $\bullet$}}
\begin{document}
\draft

\title{Dissipative Transport in Quantum Hall Ferromagnets by Spinwave 
Scattering}
\author{A. G. Green and N. R. Cooper}
\address{Theory of Condensed Matter Group, Cavendish Laboratory, 
Madingley Road, Cambridge, CB3 0HE,
United Kingdom.}
\date{\today}

\maketitle

\begin{abstract}

We report on a study of the effect upon electrical transport of 
spinwave scattering from charged quasiparticles in $\nu=1$ quantum 
Hall ferromagnets (QHFs), including both Heisenberg (single layer)
and easy-plane (bilayer) cases. We derive a quantum Langevin equation 
to describe the 
resulting diffusive motion of the charged particle and use this to 
calculate the contribution to low temperature conductivity from
a density of charged particles. This conductivity has a power 
law dependence upon temperature. The contribution is small at low 
temperatures increasing to a large value at relatively modest temperatures.
We comment upon high temperature transport and upon the contribution
of scattering to the width of the zero bias peak in inter-layer tunneling 
conductance.

\end{abstract}


\narrowtext

\section{Introduction}

In quantum Hall ferromagnets (QHFs)\cite{QHEbook} there are two
important energy gaps: the spinwave gap, describing the minimum energy
spinwave excitation; and the quasiparticle gap, describing the energy
of a widely-separated quasiparticle/quasihole pair. The spinwave gap
is due to the Zeeman energy in the single layer, real spin QHF and to
the inter-layer tunnelling in the bilayer pseudo-spin QHF. The
quasiparticle gap is determined ultimately by the Coulomb
energy. These energies are independent and widely separated. There is,
therefore, a regime of temperatures between these two energies, where
a large population of neutral spin-waves exists, yet where charged
quasi-particles are rather dilute. 

These spin-waves have a direct
effect upon a number of experimental observables. The temperature
dependence of magnetisation and nuclear relaxation rates have been
studied both experimentally and
theoretically\cite{Barrett95,Manfra96,Read95,Haussmann97,Timm98,Kasner00}.
In addition to these magnetic properties, one can expect {\it
transport} properties of the system to be affected by the
thermally-excited spin-waves.
The heat flow carried by the spin-waves will lead to a thermal
conductivity that follows a power-law of temperature (rather than
being exponentially suppressed as in a standard quantum Hall state).
One also expects a power-law contribution to the diffusion
thermopower, since for weak disorder this a measure of the entropy per
particle\cite{chr}.
In the present paper we consider the consequences of a large thermal
population of spin-waves on the electrical transport properties,
motivated in part by the observation that a surprisingly low temperature 
is required for a good quantum Hall effect in the 
QHFs\cite{Schmeller95,SpielmanEPW00}.  

Much of the truly novel physics of QHFs stems from the nature of the
underlying quantum Hall state. In particular, spin and charge
fluctuations are intimately linked so that the magnetic vorticity and
charge density are proportional to one another\cite{sondhi}. One
consequence of this relationship is that spin-waves, while electrically
neutral, carry a dipole moment and thus interact electrostatically
with any charged excitations. 
Here, we study how the scattering of
spin-waves off charged excitations, assumed present either by activation
or by slight departures from $\nu=1$, affects the diffusion of the
charged excitations. Can it lead to a significant, finite-temperature
 enhancement of the longitudinal conductivity of these quasiparticles? 
We shall consider 
both Heisenberg and easy-plane
QHFs at $\nu=1$, relevant for single layer and bilayer quantum Hall
systems, respectively.

The scattering of spinwaves from charged quasiparticles in the QHF
has been considered previously in Ref.\cite{Kasner00}. The focus 
in that work was on the temperature dependence of magnetisation and 
spectral properties of the electronic Green's function and its effect
upon tunnelling conductance. In contrast, here we consider the 
consequences of quasi-particle/spinwave scattering upon in-plane
transport. 

Section~\ref{sec:diff} contains the principal results of the paper.
We begin by describing the model we use, and in \ref{sec:simple}
provide a simple derivation of the quasiparticle diffusion constant in
terms of a force-force correlation function.
Sections~\ref{sec:collcoord} and~\ref{sec:fv} provide a systematic
derivation of this result, based upon the use of collective co-ordinates
and an influence functional.  In section~\ref{sec:lowt}
we use the formula derived to calculate the diffusion
constants for Heisenberg and easy-plane QHFs at low
temperature. The results are discussed in section~\ref{sec:disc}. 
We find that the contribution to longitudinal conductivity is small at
temperatures much less than the spin-stiffness. The temperature 
dependence is strong,  particularly in the bilayer QHF. We present 
arguments why the conductivity may be expected to become large at 
relatively modest temperatures. In
\ref{sec:tunnel} we comment on the effects of the spin-wave scattering
in our model on the tunnelling conductance of a bilayer system QHF.

\section{Quasi-particle Diffusion Constant}

\label{sec:diff}

We shall study the longitudinal conductivity of disorder-free QHFs at
$\nu=1$ containing a dilute gas of charged excitations. These may be
present due either to thermal activation or to (local) deviations of
density that cause (local) departures from precisely $\nu=1$.  The
diffusion of these mobile charges will determine the longitudinal
conductivity, $\sigma_{xx}$.  Since we treat the charges as
independent, positive and negative charges will contribute in equal
measure to $\sigma_{xx}$ (we consider the strong field limit of the
lowest Landau level, for which there is a particle-hole symmetry at
$\nu=1$).  For simplicity, we represent the total concentration of
mobile charges (be they positive or negative) in terms of a single
filling fraction $\delta\nu$, such that the number density of charges
is $\delta\nu \bar\rho$, where $\bar\rho \equiv eB/h$ is the density
of states in a Landau level.  We shall also refer to these mobile
charges as ``quasi-particles'' -- independent of their internal spin
structure -- except when it is important to make a distinction.

What is the motion of a quasi-particle under an applied electrical
field?  At zero temperature, the quasi-particle moves perpendicular to
the applied electric field and contributes to the Hall conductivity,
but not to the longitudinal conductivity. This may be appreciated on
various grounds.  Firstly, translation invariance allows the electric
field to be removed by a Lorentz transformation to a reference frame
moving perpendicular to the electric field. In this frame the
quasi-particle will be stationary. Its motion in the lab frame is,
therefore, perpendicular to the electric field. A second way to
appreciate this motion, and one that will prove useful in
understanding the processes that we consider here, is to note that the
kinetic energy of a particle in the lowest Landau level is
quenched. In moving parallel to the electrical field, the
quasi-particle would absorb energy from this field. Since there are no
states in the lowest Landau level that have different energy, the
quasi-particle is constrained energetically to move on contours of
equipotential.

At a finite temperature, the quasi-particle moves in a heat bath of
spin-waves.  The heat bath defines a rest frame; translational
invariance is broken and the longitudinal conductivity is not
zero. [We assume that the spin-waves are in thermal
equilibrium in the rest frame.  Ultimately this is due to
equilibration of spin-waves with the lattice by interaction with
phonons.]  The scattering of spin-waves induces a diffusive motion of
the quasi-particle.  It provides a mechanism by which the
quasi-particle may lose energy to spin-waves and so move down an
applied potential gradient.
 
\subsection{Simplified derivation}

\label{sec:simple}

We first provide a simple calculation of the diffusion constant, which
illustrates how spin-wave scattering leads to quasi-particle
diffusion.  We treat the spin-waves as free particles in the absence
of the quasi-particle, and study the scattering of these modes from
the perturbing quasi-particle. This scattering induces motions of the
quasi-particle.  Were we to treat the spin-waves as the linearised
excitations in the presence of the quasi-particle, there would be no
scattering and hence no motions of the quasi-particle. Although we
offer here no formal derivation of the approach we use, we expect it
to capture the nonlinearities that arise from the fact that the
displacement of the particle cannot be treated as a small fluctuation,
and thus the spin-waves cannot be viewed as decoupled quadratic
fluctuations. This is confirmed by the agreement of the results of
this approach with the systematic derivations of
sections~\ref{sec:collcoord} and~\ref{sec:fv}.
  
The first step in our calculation is to write down the rate of a
process where a single spin-wave scatters off a quasi-particle:
\begin{eqnarray}
\Gamma_{\delta R,i,f}
 &=&  
\frac{2 \pi}{\hbar}
\left| \langle f |\Delta \hat {\cal H} | i \rangle \right|^2 
\delta \left( E^0_f-E^0_i\right)
\nonumber\\
& &
\;\;\;\;\;\;\;\;\;\;\;\;\;\;\;
\times
\delta \left[ {\delta\bbox{R}} -\ell^2  
\hat{\bbox{z}} \times (\bbox{k}^f-\bbox{k}^i)/\hbar\right]
\label{Scattering_rate}
\end{eqnarray}
In this expression, $|i\rangle,|f\rangle$ are initial and final states
of the (unperturbed) spin-wave system with total momenta
$\bbox{k}^{i,f}$ and total energies $E^0_{i,f}$, which are coupled by
the perturbation $\Delta \hat {\cal H}$ arising from the presence of a
quasi-particle, whose position is displaced by $\delta \bbox{R}$ under
the scattering. $\ell\equiv\sqrt{\hbar/eB}$ is the magnetic length.
The second delta function embodies the important physics of the lowest
Landau level.  The two components of position in the lowest Landau
level are conjugate to one another. A change in the momentum of the
quasi-particle, $\delta \bbox{k}$, is equivalent to a change
$\ell^2 (\hat{\bbox{z}} \times \delta\bbox{k})/\hbar$ in its
position. The rate of diffusion of the particle arising from this
scattering process is obtained from Eq.(\ref{Scattering_rate}) by
averaging $|\delta\bbox{R}|^2$ over a thermal distribution of initial
spin-wave states. This determines the rate of increase of the mean
square displacement of the quasi-particle, and hence the diffusion
constant $D\equiv \frac{1}{4} d \langle
|\bbox{R}(t)-\bbox{R}(0)|^2\rangle/dt$:
\begin{equation}
D = 
\frac{\pi \ell^4}{2\hbar^3}\sum_{if} 
\rho^0_i
\left| \langle
f | \Delta \hat {\cal H} | i 
\rangle \right|^2 |\bbox{k}^i-\bbox{k}^f|^2
\delta \left( E^0_f-E^0_i \right),
\label{eq:diffusion}
\end{equation}
where $\rho^0_i = e^{-E^0_i/k_B T}/Z$.
The contribution to conductivity from a dilute (non-degenerate) gas of
such quasi-particles may be deduced from Eq.(\ref{eq:diffusion}) using
the Einstein relation;
\begin{equation}
\sigma_{xx}  =  e^2 D \frac{d n}{d \mu} 
   =  \frac{e^2 D \delta\nu \bar\rho }{k_B T}.
\label{eq:einstein}
\end{equation}
$\delta\nu \bar\rho$ is the average number density of quasi-particles.
Eq.(\ref{eq:diffusion}) and the resulting expression for the conductivity
may be rewritten in terms of a force-force correlation function as follows: 
\begin{eqnarray}
\sigma_{xx}&=&\delta \nu
\frac{e^2}{h}
\nonumber\\
& \times&
\lim_{\omega \rightarrow 0}
\left.
\frac{
{\cal I}m
\langle
[\hat{\bbox{k}}, \Delta \hat {\cal H}](\tilde \omega)
\cdot [\hat{\bbox{k}}, \Delta \hat {\cal H}](-\tilde \omega)
\rangle
}{4 \pi \bar{\rho} \hbar^4 \omega}
\right|_{i \tilde \omega \rightarrow \omega + i \delta},
\label{conductivity1}
\end{eqnarray}
where we have used the fluctuation dissipation relation to express
our result in terms of a retarded correlation function.
The diffusion and conductivity of the quasi-particle are related to
the forces exerted upon the quasi-particle by spin-waves from the 
heat-bath. Before going on to calculate Eq.(\ref{conductivity1}) in
various experimental regimes, we first give a more rigorous derivation
using the collective coordinate technique.

\subsection{Collective Coordinates}
\label{sec:collcoord}

The collective coordinate technique\cite{rajaraman} provides a systematic way
of obtaining an effective theory for the interaction of a Skyrmion
with the heat bath of spin-waves. This method has been fruitfully 
applied to the study of polaron transport in Ref.\onlinecite{Castro92}. We
follow the methods of this paper quite closely. The starting point
is the sigma-model effective action for the QHF\cite{sondhi};
\begin{eqnarray}
{\cal S}
&=&
\int dt d\bbox{r}
\left[
\frac{\bar \rho}{2} \bbox{A}[\bbox{n}] \cdot \partial_t \bbox{n}
- 
\frac{\rho_s}{2} (\nabla \bbox{n})^2
+
\bar \rho g n_z
\right]
\nonumber\\
& &-
\int dt \;V[\rho].
\label{sigma_model}
\end{eqnarray}
$\bbox{n}$ is an $O(3)$ vector field giving a coherent-state
representation of the spin.  The first term in this action is the
Berry phase term describing the spin dynamics. It embodies the
commutation relations of the spin operators.  The second and third
terms describe the exchange and Zeeman energy of the QHF ($\rho_s$ is
the spin stiffness and $g$ the Zeeman energy per electron). 
For simplicity, we choose to study
only the case of an Heisenberg ferromagnet in this section; 
the approach we use can
easily be adapted for the easy-plane case.

It is the final term in Eq.(\ref{sigma_model}) that distinguishes the
QHF from a conventional ferromagnet; it describes the identity between
charge and magnetic vorticity\cite{sondhi} discussed in the
introduction. The charge density associated with a spin distortion,
for $\nu=1$, is given by
\begin{equation}
\rho
=
\frac{e}{8 \pi} \epsilon_{ij}
\bbox{n}\cdot \partial_i \bbox{n} \times \partial_j \bbox{n}.
\label{top_density}
\end{equation}
The final term in Eq.(\ref{sigma_model}) indicates the Coulomb self
interaction of the spin field. The foremost consequence of
Eq.(\ref{top_density}) is that magnetic vortices or Skyrmions in the
QHF carry unit charge. The static Skyrmion distribution,
$\bbox{n}_0(\bbox{r})$, is found by minimising the energy
(minus the time independent part of Eq.(\ref{sigma_model})) in
the single Skyrmion sector. This analysis was carried out in
Ref.\onlinecite{sondhi}.  We do not require any details here except
for the existence of $\bbox{n}_0$. The next step is to expand in small
fluctuations about the Skyrmion groundstate. Care must be taken with
this expansion.  If it is carried out for a static Skyrmion, some of
the normal modes are found to have zero energy. They correspond to
translation and rotation of the Skyrmion spin distribution. We use the
collective coordinate technique in order to handle these zero
modes. The basic idea is to exclude the zero modes from the spin-wave
field and to elevate the Skyrmion position -- its collective
coordinates -- to be a dynamical variable, $\bbox{R}(t)$. The
spin-wave expansion about the moving Skyrmion spin distribution is
given by
\begin{eqnarray}
\bbox{n}(\bbox{r},t)
&=&
\bbox{n}_0 (\bbox{r}-\bbox{R}(t)) 
\sqrt{ 1 - |\bbox{l}(\bbox{r}-\bbox{R}(t),t)|^2} 
\nonumber\\
& &+ \bbox{l}(\bbox{r}-\bbox{R}(t),t).
\label{spin-wave_expansion}
\end{eqnarray}
$\bbox{l}(\bbox{r}-\bbox{R}(t),t)$ is the spin-wave field in the presence
of the Skyrmion at the point $\bbox{R}(t)$. It may be expanded in terms of
spin-wave eigenfunctions as follows:
\begin{equation}
l(\bbox{r}-\bbox{R}(t),t)
=
\sum_{n=1}^{\infty} q_n(t) \eta_n(\bbox{r}-\bbox{R}(t),t),
\label{collective_coordinate_expansion}
\end{equation}
where $\eta_n(\bbox{r}-\bbox{R}(t),t)$ is a spin-wave eigenstate in the 
presence of the Skyrmion and $q_n(t)$ is a time dependent occupation 
of this mode. Since the Skyrmion spin distribution changes in time as
the Skyrmion moves, the eigenmodes themselves change. 
It is this additional time dependence that induces transitions
between the spin-wave eigenmodes; although 
$\langle n(t) |m(t) \rangle =0$ for $m \ne n$, 
$\langle n(t) |m(t+\delta t) \rangle  \ne0$ allowing transitions between 
them. These effects are encoded in the Berry phase term of 
Eq.(\ref{sigma_model}). Upon substituting the collective coordinate 
expansion, Eq.(\ref{collective_coordinate_expansion}), into the 
first term of Eq.(\ref{sigma_model}), we find
\begin{eqnarray}
\lefteqn{
\int dt d\bbox{r}
\frac{\bar \rho}{2} \bbox{A}[\bbox{n}]\cdot \partial_t \bbox{n}
=
\int \!dt 
\;\hbar \pi \bar \rho  \hat{\bbox{z}} \cdot \bbox{R} \times \dot{\bbox{R}}}
\nonumber\\
& &
+
\int dt d\bbox{r}
\left[
\frac{\hbar \bar \rho}{4}
\dot{\bbox{R}}\cdot i\bar l \nabla l
+
i\frac{\hbar\bar \rho}{4} l \partial_t l
\right]
\label{Berry_phase}
\end{eqnarray}
We have adopted the complex notation $l=l_1+i l_2$, $\bar l=l_1-il_2$.
The first term describes the bare dynamics of the Skyrmion\cite{stoneskyrmion}.
It is the usual action for a particle with Magnus force dynamics. 
The third term 
describes the spin-wave dynamics. The second term describes the 
time dependence of the spin-wave field arising from the motion of
the Skyrmion ($d_t l = \partial_t l - \dot{\bbox{R}} \cdot \nabla l$). 
It is this term that gives rise to the non-orthogonality of
spin-wave eigenstates at different times and permits scattering
between them. Notice that it consists of the coupling of the 
Skyrmion velocity to the total spin-wave momentum;
\begin{equation}
k_j
=
 i\frac{\hbar \bar \rho}{4}
\int\!\! d\bbox{r} \;\bar l \nabla_j l.
\label{spin-wave_momentum}
\end{equation}
The remaining terms in the joint spin-wave/Skyrmion effective action
are obtained by substituting
Eq.(\ref{collective_coordinate_expansion}) into the time independent
part of Eq.(\ref{sigma_model}). The resulting expressions include
terms describing the exchange and Zeeman energies of the spin-wave
distortion and terms describing the interaction of the spin-waves with
the Skyrmion. This interaction may be divided into two parts; exchange
interactions and Coulomb interactions. The exchange interactions are
local in space, whereas the Coulomb interactions are spatially
non-local, due to the long range of the Coulomb potential. 
We neglect local interactions in our treatment of the interaction of 
quasi-particles with spin-waves. This approximation 
is justified provided 
the typical spin-wave wavelength is large compared to the size of the
Skyrmion, in which limit the exchange interaction is suppressed
relative to the non-local Coulomb interaction. Retaining only Coulomb
coupling, and adding the spin-wave energy to Eq.(\ref{Berry_phase}), we
find the following joint spin-wave/Skyrmion effective action:
\begin{eqnarray}
\lefteqn{
{\cal S}[\bbox{R},l, \bar l]
=
\hbar
\pi \bar \rho \int dt \hat{\bbox{z}} \cdot \bbox{R} \times \dot{\bbox{R}}
+
\frac{\hbar \bar \rho}{4}
\int dt d\bbox{r}
\dot{\bbox{R}}\cdot i\bar l \nabla l}
& &
\nonumber\\
& &
+
\frac{1}{2}
\int dt d\bbox{r}
\bar l
\left[
i\frac{\hbar \bar \rho}{2} \partial_t
-
\rho_s \nabla^2
-
\bar \rho g
\right]
l
\nonumber\\
& &
-
\int dt d\bbox{r} d\bbox{r}'
V(\bbox{r}-\bbox{r}') 
\rho_{\bbox{n}_0}(\bbox{r}-\bbox{R}(t)) 
\rho_{\bbox{l}}(\bbox{r}'-\bbox{R}(t),t),
\label{effective_action}
\end{eqnarray}
where $\rho_{\bbox{n}_0}$ is the charge density of the Skyrmion and
$\rho_{\bbox{l}}$ is the charge density associated with spin-waves.
$\rho_{\bbox{l}}$ is given by
\begin{equation}
\rho_{\bbox{l}}
=
-i\frac{e}{8 \pi}
\epsilon_{ij} \partial_i \bar l \partial_j l.
\label{spin-wave_charge_density}
\end{equation}
Before proceeding to calculate the Skyrmion dynamics from
Eq.(\ref{effective_action}), let us make a few comments about the
comparison of the Skyrmion/spin-wave problem with the polaron/phonon
problem\cite{Castro92}.  The dynamics of spins is entirely determined by
their commutation relations and, importantly, there is no kinetic term
in their Hamiltonian. This results in different dynamics for the 
Skyrmion and polaron. In
the former case, one finds Magnus-force dynamics describing the motion
of a Skyrmion perpendicular to an applied force. In the latter case,
however, the dynamics have a conventional ballistic form. The second
consequence is the absence of (multiple) spin-wave Cherenkov 
processes. Such terms are found in the phonon/polaron case through the
collective coordinate expansion of the kinetic terms in the
Hamiltonian. They are forbidden by energy conservation in the Skyrmion
case, unless one allows for internal modes of the
Skyrmion\cite{FertigBCM96} (which we neglect here, under the
assumption that the drift velocity of the Skyrmion is less than the
critical velocity derived in Ref.\onlinecite{FertigBCM96}). These
facts were missed in a previous analysis of the Skyrmion problem by
Villares Ferrer and Caldeira\cite{Villares00}.  Despite these key
differences, when Cherenkov processes are neglected, we find that the
coherent state representation of the Skyrmion and polaron problems are
very similar and that the damping and diffusion of Skyrmions is very
similar to that of polarons.

\subsection{Feynman-Vernon Influence Functional}
\label{sec:fv}

Our goal in this subsection is to use the Skyrmion/spin-wave effective
action, Eq.(\ref{effective_action}), to study the Skyrmion
dynamics in the presence of the heat bath of spin-waves. In order to
carry out this analysis one may use the Feynman-Vernon influence
functional approach\cite{Feynman63,Schmid82}.  The application 
of this approach to the present problem is very similar to its application in
the polaron case \cite{Castro92}. The calculation proceeds through a number
of steps, but the basic idea is the following: the reduced density
matrix for the Skyrmion is found by tracing the total system density
matrix over the spin-wave degrees of freedom, {\it i.e.} by
`integrating out' the spin-waves.  The time evolution of this density
matrix may be expressed in terms of a superpropagator, which is in
turn expressed in terms of influence functionals that encode the
effect of the spin-waves on the Skyrmion propagation. The result
of such a calculation is to express the damping and diffusion of
Skyrmions in terms of momentum-momentum correlation functions of the
spin-wave heat bath in the presence of the Skyrmion potential.  Full
details of such a calculation may be found in Ref.\onlinecite{Castro92}.  
A brief summary of an equivalent calculation using Keldysh 
techniques\cite{Keldysh64} is given in Appendix\ref{app}. 
The main approximation in
carrying out this procedure is that the Skyrmion is displaced by a 
distance less than the spinwave wavelength at each scattering process.

The above analysis results in the following Langevin equation describing
the motion of the Skyrmion in the presence of the spin-wave heat-bath:
\begin{eqnarray}
2 \pi \hbar \bar \rho
\hat{\bbox{z}} \times \dot{\bbox{R}}
+
2 \bar \gamma \dot{\bbox{R}} -e \bbox{E}
=
\bbox{\zeta}(t)
\nonumber\\
\langle \zeta_i(t) \zeta_j(t') \rangle
=
2 \overline D \delta_{ij} \delta(t-t').
\label{Langevin_equation}
\end{eqnarray}
The first term in this equation describes the transverse motion of 
the Skyrmion in response to an applied force. The second term describes 
dissipation of the Skyrmion motion due to the scattering of spin-waves.
$\bbox{E}$ is an applied electric field. The term on the right hand side
describes diffusive motion of the Skyrmion due to the scattering of 
spin-waves. The dissipation and diffusion constants are related to the 
spin-wave momentum-momentum correlator {\it via}
\begin{eqnarray}
\bar \gamma
&=&
\lim_{\omega \rightarrow 0}
\omega {\cal I}m \Gamma(\omega)
\nonumber\\
\overline D
&=&
2 \overline \gamma T
\nonumber\\
\Gamma(t)
&=&
-i \frac{\theta(t)}{4 \hbar} 
\langle [
\hat k_i(t), \hat k_i(0)
] \rangle.
\label{gamma_and_D}
\end{eqnarray}
These correlation functions account for both thermal and zero-point
fluctuations of the background spin-field. Although zero-point
fluctuations make important contributions to the renormalization of the
Skyrmion polarization and energy\cite{Abolfath98,Walliser00},
our results show that
they do not contribute significantly to dissipation in either the single
layer Heisenberg QHF or the ordered phase of the bilayer, easy-plane QHF.

The contribution of a dilute gas of Skyrmions with number density
$\delta \nu \bar \rho$ to the longitudinal conductivity may be 
deduced from Eq.(\ref{Langevin_equation}). In the limit of 
$2 \bar \gamma / h \bar \rho \ll 1$ it is given by
\begin{equation}
\sigma_{xx}
=
\delta \nu \frac{e^2}{h}
\frac{\bar \gamma}{\pi \hbar \bar \rho}.
\label{conductivity2}
\end{equation}
This result may also be derived from the Einstein
relation, Eq.(\ref{eq:einstein}). The appropriate diffusion constant
for use in Eq.(\ref{eq:einstein}) is $D= \overline D/(2 \pi \hbar \bar
\rho)^2$. This can be seen by using Eq.(\ref{Langevin_equation}) to
calculate $\langle |\bbox{R}(t)-\bbox{R}(0)|^2 \rangle =4Dt =4\overline
Dt/(2 \pi \hbar \bar \rho)^2$.

Eq.(\ref{conductivity2}) represents the Skyrmion conductivity in 
terms of the 
spin-wave momentum-momentum correlation function in the presence of
a static Skyrmion, Eq.(\ref{gamma_and_D}). This is our primary result.
We have used the 
very general Feynman-Vernon/Keldysh techniques in order to derive 
this result, however, it may also be obtained quite straightforwardly 
from the effective action Eq.(\ref{effective_action}) using the 
Kubo formula \cite{Kubo66}. After expanding 
the momentum-momentum correlation
function over spin-wave states in the presence of the Skyrmion
and then expressing these states in terms of free spin-wave states
using the lowest order perturbation theory, Eq.(\ref{conductivity2})
may be expressed in terms of an average over free spin-waves:
\begin{eqnarray}
\sigma_{xx}
&=&
\delta \nu
\frac{e^2}{h}
\nonumber\\
& \times &
\lim_{\omega \rightarrow 0}
\left.
\frac{
{\cal I}m
\langle
[\hat{\bbox{k}}, \Delta \hat {\cal H}_{\bbox{R}}](\tilde \omega)\cdot
[\hat{\bbox{k}}, \Delta \hat {\cal H}_{\bbox{R}}](-\tilde \omega)
\rangle
}{
4 \pi \bar \rho \hbar^4 \omega}
\right|_{i \tilde \omega \rightarrow \omega + i \delta}\!\!\!\!\!\!\!\!\! .
\label{conductivity4}
\end{eqnarray}
This recovers the result obtained in~\ref{sec:simple} by simple 
Fermi's Golden rule arguments. There were two key approximations
in the derivation of Eq.(\ref{conductivity4}). 
The first is 
the requirement that the recoil of the Skyrmion after any particular
scattering event is much less than the wavelength of the 
spin-waves involved. Secondly, we have expanded perturbatively in 
the interaction between spin-waves and the Skyrmion. This is 
equivalent to the Born approximation for the scattering of spin-waves.
Notice that we have made no assumptions about the nature of the 
interaction between the spin-waves and Skyrmion in our derivation.

\subsection{Low temperature conductivity.}
\label{sec:lowt}

We are now in a position to calculate the contribution to conductivity 
from spin-wave scattering. First we consider the {\it Heisenberg} case,
for which the long-wavelength spin-wave dispersion is
\begin{equation}
E^{0}(k) = g + 2\rho_sk^2/\bar\rho .
\end{equation}
At the lowest temperatures, the dominant interaction between
spin-waves and charged excitations is through the Coulomb
interaction. The perturbation in the spin-wave Hamiltonian due to the
presence of a Skyrmion at point $\bbox{R}$ is given by
\begin{equation}
\Delta \hat {\cal H}_{\bbox{R}}= \int d\bbox{r} V(\bbox{R}-\bbox{r}) 
\rho_{\bbox{l}}(\bbox{r},t),
\label{potential_interaction}
\end{equation}
where the Skyrmion has been treated as a point charge on the
lengthscale of the scattered spin-waves. The commutator of this
Hamiltonian with the spin-wave momentum operator is given by $[\hat
k_i, \Delta \hat {\cal H}_{\bbox{R}}] = i \hbar \partial_{\bbox{R}}
\Delta \hat {\cal H}_{\bbox{R}}$, where we have used translational
invariance to express in terms of the derivative with respect to
the Skyrmion co-ordinate. Substituting this into Eq.(\ref{conductivity4})
and using the Coulomb interaction potential, $V(\bbox{q})=e^2/2
\epsilon |\bbox{q}|$, the conductivity of a dilute Skyrmion gas may be
expressed in terms of a correlation function of the free spin-wave
topological density;
\begin{eqnarray}
\sigma_{xx}
&=&
\delta \nu \frac{e^2}{h} 
\frac{e^2}{16 \pi \epsilon^2 \hbar^2 \bar \rho}
\nonumber\\
& \times &
\lim_{\omega \rightarrow 0}
\int \frac{d \bbox{q}}{(2 \pi)^2}
\left.
\frac{
{\cal I}m
\langle 
\rho_{\bbox{l}}(\bbox{q},\tilde \omega)
\rho_{\bbox{l}}(-\bbox{q},-\tilde \omega)
\rangle
}{\hbar \omega}
\right|_{i \tilde \omega +i\delta}.
\label{conductivity5}
\end{eqnarray}
Calculating with the free spin-wave part of the effective action,
Eq.(\ref{effective_action}), and the spin-wave charge density,
Eq.(\ref{spin-wave_charge_density}), we find
\begin{equation}
\sigma_{xx}=\delta \nu \frac{e^2}{h}
\frac{1}{6 \times 2^{11} \pi} \frac{E^2_C (k_BT)^2}{\rho_s^4}.
\label{easy_axis_conductivity}
\end{equation}
at temperatures above the Zeeman gap and exponential suppression
with a factor $e^{-2g/k_B T}$ at temperatures below the
gap. $E_C=e^2/4 \pi \epsilon \ell$ is the characteristic Coulomb energy.

The case of the {\it easy-plane} pseudo-spin ferromagnet is a little 
more subtle. The effective action in this case is obtained by 
replacing the Zeeman term, $\bar \rho g n_z$, in Eq.(\ref{sigma_model}) by an
easy-plane anisotropy or capacitance energy, $\gamma n_z^2$. 
The pseudo-spin lies in the plane in the groundstate and the
topological defects are vortices of the in-plane pseudo-spin
orientation. The cores of these vortices are non-singular due to the
pseudo-spin rising up or below the xy-plane.  Depending upon the
vorticity and orientation in the core, these vortices may carry $\pm
1/2$ charge in addition to their $\pm$ vorticity.  These charged
vortices are known as merons\cite{Yang94,QHEbook}. The exchange interaction
energy between vortices varies logarithmically with their
separation. At low temperatures this binds vortices into pairs of
opposite vorticity. The charge carriers at low temperature are,
therefore, bound pairs of merons with charge $\pm 1$\cite{QHEbook}. These bound
pairs behave as Skyrmions in the Heisenberg case.
On lengthscales large compared with the meron
separation, the meron pair may be viewed as a point charge. At low
temperatures, therefore, we may use the interaction,
Eq.(\ref{potential_interaction}), to model the scattering of
spin-waves and Eq.(\ref{conductivity5}) to calculate the conductivity.
The calculation is a little different to that of the Heisenberg
spin-waves.  With an easy-plane anisotropy, $\gamma n_z^2$, the
effective action Eq.(\ref{sigma_model}) has a spin-wave dispersion 
given by\cite{Fertig89}
\begin{equation}
E^{0}(k) = 2 \sqrt{ \rho_s \bbox{k}^2 (\rho_s \bbox{k}^2 + 2 \gamma)}/\bar \rho
. 
\label{dispersion}
\end{equation}
This dispersion is linear at low momentum crossing over to a quadratic
behaviour at a momentum $\bbox{k}_c=\sqrt{2 \gamma/ \rho_s}$
An effective theory for the linearly dispersing
modes may be obtained in terms of the in-plane orientation of the
pseudo-spin field by integrating out $n_z$ from the effective action,
Eq.(\ref{sigma_model}). The result is a quantum XY-model;
\begin{equation}
{\cal S}
=
\int dt d\bbox{r}
\frac{\rho_s}{2}
\left[
\dot \phi^2 v^{-2} - (\nabla \phi)^2
\right],
\label{easy_plane_action}
\end{equation}
where $v= \sqrt{8 \gamma \rho_s}/(\hbar \bar \rho)$ is the velocity of the 
linearly dispersing modes. The spin-wave charge density may also be expressed
in terms of $\phi$. It is given by
\begin{equation}
\rho_{\phi}
=
i \frac{\hbar \bar \rho}{16 \pi \gamma} \epsilon_{ij}
\partial_i \phi \partial_j \dot \phi.
\label{easy_plane_charge_density}
\end{equation}
The interaction between the pseudo-spinwaves and the meron pair is given by 
Eq.(\ref{potential_interaction}), replacing $\rho_{\bbox{l}}$ with $\rho_{\phi}$. The conductivity 
is given by Eq.(\ref{conductivity5}) with a similar replacement.
Calculating the conductivity at low temperatures using 
Eqs.(\ref{easy_plane_action}) and (\ref{easy_plane_charge_density}), we find
\begin{equation}
\sigma_{xx}
=
\delta \nu \frac{e^2}{h}
\frac{\pi^3}{84}
\frac{E_C^2 (k_BT)^6}{(\hbar v)^8 \bar \rho^4}.
\label{easy_plane_conductivity}
\end{equation}
This is suppressed by a factor of $(\pi \bar \rho k_B T/\gamma)^4/28$
relative to the Heisenberg ferromagnet. Turning on the anisotropy has
stiffened up the low momentum spin-waves so that fewer are thermally
excited at low temperatures leading to a corresponding reduction in
quasi-particle scattering and conductivity. At temperatures above
$k_B T\approx \gamma/\bar \rho$, significant numbers of thermally excited
spin-waves are in the quadratic part of the dispersion, Eq.(\ref{dispersion}).
These spinwaves also have sufficiently short wavelength to probe the 
structure of the meron pair. The conductivity is expected to 
cross over to the form given by Eq.(\ref{easy_axis_conductivity}), with a
modified pre-factor. However, the temperature $k_B T\approx \gamma/\bar \rho$
is typically rather large and the system is likely to undergo a 
Kosterlitz-Thouless
transition before this cross-over becomes apparent.
Note that for a bilayer QHF at $\nu=1$, at temperatures larger than
both the tunnelling gap and the Zeeman energy there will be scattering
of {\it both} easy-plane pseudo-spin waves, and Heisenberg ``real''
spin-waves. The present discussion indicates that in the low
temperature regime, the scattering of the ``real'' spin-waves will
give the dominant contribution to quasiparticle diffusion.

\subsection{Discussion}

\label{sec:disc}

The diffusion constants we calculate are strongly increasing functions
of temperature.  However, even for reasonably high temperatures,
within the range of applicability of the spin-wave expansion, the
diffusion constants
(\ref{easy_axis_conductivity},\ref{easy_plane_conductivity}) remain rather
small. As an illustration, we consider the Heisenberg case, with
typical parameters of $B=4\mbox{T}$ and a temperature $T=3\mbox{K}$
that is comparable to $\rho_s$ (for a narrow 2DEG with
$\epsilon_r=12.5$).  The above formula results in a longitudinal
conductivity of only $0.06 \delta\nu (e^2/h)$ for a concentration of
$\delta\nu$ quasi-particles (the conductivity for easy-plane
anisotropy is always smaller than that for the Heisenberg magnet).  It
may be difficult to observe this intrinsic diffusion owing to the
effects of disorder.

Disorder can have a dramatic effect on quasi-particle diffusion, even
if the rms disorder potential $\phi_{rms}$ is much smaller than
temperature, $e\phi_{rms}\ll k_BT$. The additional $\bbox{E}\times \bbox{B}$ drift
that the disordered potential introduces to the classical dynamics of
the quasi-particle leads to\cite{isichenko} an effective diffusion
constant $D^*$ that is enhanced over the intrinsic diffusion $D$. The
extent of this enhancement depends on the ratio $P=\phi/(BD)$.  The
intrinsic diffusion dominates ($D^*\simeq D$) provided $P\lesssim
1$. For the above parameters, this sets an upper limit of
$e\phi_{rms}/k_B \lesssim 0.17K$ on the disorder strength. For
stronger disorder, $P\gg 1$, the effective diffusion constant is
enhanced, $D^* \sim D P^{10/13}$.\cite{isichenko}
At temperatures much less than the disorder strength, $k_BT\ll e\phi$,
the transport mechanism of quasi-particles will involve thermal
activation or variable range hopping\cite{Polyakov93}.

Even in the absence of disorder, we may ask what is the conductivity 
at high temperatures when the density of thermally generated charges 
is large and the assumption of independent quasi-particles used above
breaks down. At a temperature $k_B T_{KT}= \pi \rho^R_s/2$, the
easy-plane QHF is expected to undergo a Kosterlitz-Thouless 
transition where vortices unbind due to thermal fluctuations 
($\rho_s^R$ is the thermally renormalised spin stiffness). This
has a profound effect upon the nature of transport.
Above the KT transition, the charge is carried by merons. In addition
to carrying charge in units of $\pm 1/2$, merons have a vortex
configuration of in-plane spin. The interaction between unpaired
merons and pseudo-spin-waves is, therefore, dominated by exchange.
At high temperatures, above the KT-transition, we expect exchange
scattering of thermally generated merons to lead to a conductance 
near to $\sigma = e^2/4 h$. The reason for this is the following:
the quantum XY-model displays a zero temperature phase transition 
at $v=2 \pi \rho_s/\Lambda$ ($\Lambda$ is the ultra-violet momentum cut-off.),
 where zero point fluctuations destroy long-range order.
One may use duality\cite{Cha91} at this point between the two zero 
temperature 
phases to argue that the vortex number conductivity takes a universal
value $\tilde \sigma=1/h$ at the transition point and at temperatures
above this critical point, in the quantum-critical regime. Since each
meron carries a charge $e/2$, we expect a charge conductivity
of $\sigma \approx e^2/4h$ provided that the Coulomb interaction may
be ignored.
[Of course, the conductivity $\tilde \sigma$ gives the response to a 
field that couples to vorticity and $\sigma$ the response to a field
that couples to charge. The charge and vorticity of a meron are 
independent; a meron of a particular vorticity may carry either charge.
However, the conductivity in both cases is proportional to the density 
of merons and inversely proportional to the resistance to motion of
an individual meron. We, therefore, anticipate the simple relationship
$\sigma=e^2 \tilde \sigma /4$.]
This supports the suggestion in Ref.\onlinecite{SpielmanEPW00,SpielmanEPW01} 
that a rapid increase
in longitudinal conductivity occurs at the KT-transition.

A similar analysis of the high temperature conductivity of the 
Heisenberg QHF is not possible. The Heisenberg QHF does not undergo a 
zero temperature phase transition as in the easy-plane case. The
duality arguments that lead to the prediction of a universal conductivity
in the case of the easy-plane QHF cannot be used. Recall, however, that 
in the case of low temperature response, the easy-plane QHF always has 
a lower conductivity than the corresponding Heisenberg magnet. This is
due to the stiffening of the low-energy spinwaves in the easy-plane
QHF leading to a reduction in their thermal population. Features in the 
behaviour of the easy-plane QHF due to pseudo-spin fluctuations are
expected to be stronger in the Heisenberg QHF. We expect, therefore, that
although it does not display a KT-transition, the Heisenberg QHF should show
a crossover to dramatically enhanced (and possibly universal) conductivity
at temperatures around $k_BT=\pi \rho_s^R/2$. Some circumstantial evidence 
for this is found in numerical simulations of the classical 2-dimensional 
O(3)-sigma model, where the topological compressibility is found to rise 
rapidly at around $k_B T\sim \rho_s$\cite{Blatter96}.
These considerations may explain the longstanding puzzle in the IQHE
that, although activated transport measurements at low temperatures
indicate a large gap ($\approx 4 \pi \rho_s$), one must go to much lower 
temperatures ($\approx \pi \rho_s/2$?) than the
measured gap in order to see a well formed QH state and accompanying
minimum in longitudinal conductivity\cite{Schmeller95}.

\section{Bilayer Tunnelling: Spin-Wave lifetime}

\label{sec:tunnel}

Our discussion has focused upon the transport properties of the
QHF. Of late, however, much of the focus in the study of bilayer
pseudo-spin QHFs has been on the tunnelling conductance between
layers. This conductivity shows a dramatic enhancement at zero
bias\cite{SpielmanEPW00,SpielmanEPW01}.  This is thought to be a
direct consequence of interlayer coherence and the existence of the
pseudo-spin-wave Goldstone
mode\cite{SpielmanEPW00,SpielmanEPW01,Wen92,Balents01,Stern01,Fogler01}. An
outstanding problem is to understand the height and width of the zero
bias peak in the data of
Ref.\onlinecite{SpielmanEPW00,SpielmanEPW01}. Within a perturbative
treatment\cite{Balents01,Stern01,Fogler01}, interlayer tunnelling
probes the spectral function of the pseudo-spin-waves, such that the
height and width of the zero-bias peak are set at low temperature by
the spin-wave lifetime in the limit of zero momentum. In
Refs.\onlinecite{Balents01,Stern01} it has been suggested that this
lifetime may be due to a finite density of merons. We can calculate this
lifetime within our model of spin-wave scattering off dilute isolated
quasi-particles (meron pairs) which feel no disorder, by taking
Eq.(\ref{Scattering_rate}) and integrating over final spin-wave states
and particle displacements.  The resulting scattering 
rate is\cite{Spinwave_relaxation}
\begin{equation}
\Gamma_k  =  {\delta\nu}\pi\frac{E_C^2k^3\ell^4}{\hbar^2
v}
\end{equation}
The scattering rate vanishes in the limit of small momentum. This is
because the zero momentum pseudo-spin-wave is a Goldstone mode both in
the free system and in the presence of a finite density of
quasi-particles. This scattering does not appear to be a suitable
mechanism by which the zero-bias peak may be broadened. 

At high temperatures above the KT-transition, the interaction between
pseudo-spinwaves and thermally generated merons is dominated by exchange.
The resulting $\phi \phi$-correlation function may be deduced on 
phenomenological grounds. A finite correlation length develops in the 
QC regime due to the proliferation of unbound vortices. Pseudo-spinwaves
are strongly scattered by these unbound vortices and their response 
is over-damped as a consequence. The only energy scale in the QC regime is
provided by the temperature. This sets both the correlation length,
$\xi(T) \propto T^{-1}$, and the damping rate. The resulting
pseudo-spin correlator takes the form\cite{Sachdevsbook}
\begin{eqnarray}
& &
\left.
\langle
\phi(\bbox{k}, \tilde \omega) \phi( -\bbox{k}, -\tilde \omega) 
\rangle
\right|_{i \tilde \omega \rightarrow \omega +i \delta}
\nonumber\\
&=&
\frac{1}{\rho_s}
\left( 
\bbox{k}^2+\xi(T)^{-2} -v^{-2}\omega^2 -i\alpha v^{-2}\omega T 
\right)^{-1}
\label{QC_pseudospin-wave_response}
\end{eqnarray}
for $\hbar \omega \ll kT$, where $\alpha$ is a number of order 1.
Calculating the
tunnelling current as in Refs.\onlinecite{Balents01,Stern01,Fogler01}
using the pseudo-spin-wave response function,
Eq.(\ref{QC_pseudospin-wave_response}), gives
\begin{equation}
I
\propto
\frac{2}{e^{eV/T}-1} {VT\xi(T)^4} \hbox{sign}(V).
\label{Tunnelling_Current}
\end{equation}
The Josephson singularity in differential conductance, $dI/dV$, is
not suppressed. The physics that leads
to Eq.(\ref{QC_pseudospin-wave_response}) is the scattering of 
pseudo-spin-waves from thermally excited merons. As in the case of
Coulomb scattering, this does not lead to decay of the zero-momentum 
pseudo-spinwave and so does not introduce a finite width to the 
zero-bias peak.

\section{Summary}
We have studied the scattering of spinwaves from charged excitations 
in the $\nu=1$ Heisenberg and easy-plane QHFs. This scattering leads 
to a diffusive motion of the charged quasiparticles, described by 
a quantum Langevin equation. The resulting contribution to low
temperature conductivity follows characteristic power-laws for a
given density of charge carriers. This contribution to conductivity is
small.

We have argued on the basis of duality that the conductivity of the 
easy-plane QHF at temperatures above the KT-transition crosses over 
to a universal value. Such arguments do not apply in the Heisenberg case. 
However, from a comparison of the low temperature behaviour of the 
easy-plane and Heisenberg QHF we tentatively suggest a similar crossover
at high temperatures for the Heisenberg QHF.

Finally, we have considered the dissipation of pseudo-spinwaves due
to scattering from merons in the easy-plane QHF. This scattering gives
rise to a pseudo-spinwave relaxation rate that goes to zero at zero
pseudo-spinwave momentum. It cannot, therefore, give rise to a finite 
width of the zero bias interlayer tunnelling peak.


\begin{thebibliography}{10}
\bibitem{Spinwave_relaxation}
The rate at which spinwaves equilibrate with the lattice is, therefore, 
  proportional to $T^3$, provided that there is some density of pinned charge.
  This is much greater than the quasiparticle scattering-rate, which is 
proportional to $T^7$.
\end{thebibliography}

\begin{appendix}
\section{Derivation of the Langevin Equation}
\label{app}

In Ref.\onlinecite{Castro92}, Castro Neto and Caldeira used collective 
coordinates and the Feynman-Vernon influence functional technique to
derive the reduced density matrix describing the motion of a 
polaron in the presence of a heat bath of phonons. A Langevin
equation for the polaron motion may be deduced from this density
matrix. Precisely the same method may be applied to determine the
reduced density matrix describing the motion of a Skyrmion in the 
presence of a heat bath of spinwaves. In this appendix, we use the 
alternative, but completely equivalent technique of Keldysh field
theory\cite{Keldysh64} to determine the Langevin equation describing Skyrmion motion.

The Skyrmion position and the spinwave heat bath are described
by $\bbox{x}$($\bbox{y}$) and $l_+$($l_-$) on the forward
(backwards) part of the Keldysh time contour\cite{Keldysh64}. The joint 
spinwave/Skyrmion action on the forward part of the contour is 
given by Eq.(\ref{effective_action}) with $\bbox{R} \rightarrow \bbox{x}$;
\begin{eqnarray}
& &
{\cal S}[\bbox{x},l_+, \bar l_+]
\nonumber\\
&=&
\int dt 
\left[
\hbar \pi \bar \rho \hat{\bbox{z}}\cdot \bbox{x} \times \dot{\bbox{x}}
-V(\bbox{x}) +\bbox{k}_+\cdot \dot{\bbox{x}}
\right]
+
{\cal S}_{\bbox{x}}[l_+, \bar l_+]
\end{eqnarray}
where ${\cal S}_{\bbox{x}}[l_+, \bar l_+]$ is the action for 
spinwaves in the presence of a static Skyrmion at point $\bbox{x}$
and $\bbox{k}_+=i \hbar \bar \rho \int d \bbox{x} \bar l_+ \nabla l_+/4$
is the total spinwave momentum. A similar action, 
${\cal S}[\bbox{y},l_-, \bar l_-]$ describes the motion on the return part 
of the Keldysh contour.

The next step is to make a Keldysh rotation to classical and quantum
components of the fields $l_+$, $l_-$ and the coordinates $\bbox{x}$,
$\bbox{y}$;
\begin{eqnarray}
l_{cl/q}
&=&
(l_+ \pm l_-)/2
\nonumber\\
\bbox{R}/\bbox{r}
&=&
(\bbox{x} \pm \bbox{y})/2.
\end{eqnarray}
The classical and quantum coordinates, $\bbox{R}$ and $\bbox{r}$
may be interpreted as the centre of mass of the Skyrmion wavefunction
and its spatial extent, respectively. Integrating out the spinwave 
fluctuations and retaining terms to quadratic order in $\bbox{R}$ and
$\bbox{r}$ [This is called the Born approximation in 
Ref.\onlinecite{Castro92}.
It requires that the displacement of the Skyrmion in a single scattering
process is less than the wavelength of the spinwaves involved] we
obtain the following effective Keldysh action for the Skyrmion 
coordinates:
\begin{eqnarray}
{\cal S}[\bbox{R},\bbox{r}]
&=&
\int dt 
\left[
4 \pi \hbar \bbox{r}\cdot \hat{\bbox{z}} \times \dot{\bbox{R}}
+2 \bbox{r}\cdot \nabla V(\bbox{R})
\right]
\nonumber\\
&+&
\int dt_1 dt_2
\dot{\bbox{r}}(t_1)\cdot \dot{\bbox{R}}(t_2)
\langle \bbox{k}(t_1)\cdot \bbox{k}(t_2) \rangle^R_{\bbox{R}}
\nonumber\\
&+&
\int dt_1 dt_2
\dot{\bbox{R}}(t_1)\cdot \dot{\bbox{r}}(t_2)
\langle \bbox{k}(t_1)\cdot \bbox{k}(t_2) \rangle^A_{\bbox{R}}
\nonumber\\
&+&
\int dt_1 dt_2
\dot{\bbox{r}}(t_1)\cdot \dot{\bbox{r}}(t_2)
\langle \bbox{k}(t_1)\cdot \bbox{k}(t_2) \rangle^K_{\bbox{R}}
\label{Keldysh1}
\end{eqnarray}
where $\langle \bbox{k}(t_1)\cdot \bbox{k}(t_2) \rangle^{A,R,K}_{\bbox{R}}$
are the advanced, retarded and Keldysh components of the spinwave
momentum-momentum correlator in the presence of a Skyrmion at point
$\bbox{R}$;
\begin{eqnarray}
\langle \bbox{k}(t).\bbox{k}(0) \rangle^R_{\bbox{R}}
&=&
-i\frac{\theta(t)}{ \hbar} \langle [\hat k_i(t), \hat k_i(0)] \rangle
=2\Gamma(t)
\nonumber\\
\langle \bbox{k}(t).\bbox{k}(0) \rangle^A_{\bbox{R}}
&=&
i\frac{\theta(-t)}{ \hbar} \langle [\hat k_i(t), \hat k_i(0)] \rangle
=2\Gamma(-t)
\nonumber\\
\langle \bbox{k}(t).\bbox{k}(0) \rangle^K_{\bbox{R}}
&=&
\frac{1}{ \hbar} 
\left[
\langle \hat k_i(t) \hat k_i(0) \rangle
+
\langle \hat k_i(0) \hat k_i(t) \rangle\right].
\end{eqnarray}
The Keldysh component of the spinwave momentum-momentum correlator
is related to the advanced and retarded components by a fluctuation 
dissipation relation. The potential term has been expanded for small 
$\bbox{r}$; 
$V(\bbox{R}+\bbox{r})-V(\bbox{R}-\bbox{r}) \approx 2\bbox{r}\cdot 
\nabla V(\bbox{R})$. This expansion will be justified later.

Eq.(\ref{Keldysh1}) is analogous to the effective action obtained
in Ref.\onlinecite{Castro92} using the Feynman-Vernon approach. To put 
Eq.(\ref{Keldysh1}) into the same form as that used in 
Ref.\onlinecite{Castro92},
the spinwave momentum-momentum correlation function must be expanded 
over a basis of spin-wave states in the presence of
the Skyrmion, making the identification
$g_{nm}^i
=
\bar \rho
\int d \bbox{r} \eta^*_n \nabla_i \eta_m
=
\langle n | \hat p_i | m \rangle$.

At this stage, it is convenient to rearrange some terms in 
Eq.(\ref{Keldysh1}). The second and third terms are integrated by parts 
with respect to $t_1$ and $t_2$ respectively and the fourth term
is integrated by parts with respect to both $t_1$ and $t_2$.
The result is
\begin{eqnarray}
{\cal S}[\bbox{R},\bbox{r}]
&=&
\int dt 
\left[
4 \pi \hbar \bbox{r}\cdot \hat{\bbox{z}} \times \dot{\bbox{R}}
+2 \bbox{r}\cdot V(\bbox{R})
\right]
\nonumber\\
&-&
4
\int dt_1 dt_2
\bbox{r}(t_1)\cdot \dot{\bbox{R}}(t_2)
\gamma(t_1-t_2)
\nonumber\\
&+&
i
\int dt_1 dt_2
\bbox{r}(t_1)\cdot \bbox{r}(t_2)
D(t_1-t_2)
\label{Keldysh2}
\end{eqnarray}
with
\begin{eqnarray}
2 \gamma(t)
&=&
\frac{d}{dt} \left( \Gamma(t)- \Gamma(-t) \right)
\nonumber\\
D(\omega) 
&=&
\omega \coth \left(\frac{\omega}{2T} \right) \gamma(\omega).
\end{eqnarray}
Our next few manipulations use Eq.(\ref{Keldysh2}) to derive
a Langevin equation for the Skyrmion motion.
A similar calculation is 
carried out for a simpler system in Ref.\onlinecite{Schmid82}. 
The diffusion and 
dissipation coefficients, $D(t)$ and $\gamma(t)$ are in principle non-local
in time. This implies that the Skyrmion motion may display memory effects.
We are interested in the Markovian limit where these memory effect are 
negligible. In this case $\gamma(t)= \bar \gamma \delta(t)$ and
$D(t)=\overline D \delta(t)$. Making this approximation in
Eq.(\ref{Keldysh2}) we find
\begin{eqnarray}
{\cal S}[\bbox{R},\bbox{r}]
&=&
\int
dt
2\bbox{r}\cdot
\left[
2 \pi \hbar \bar \rho \hat{\bbox{z}} \times \dot{\bbox{R}}
+
2 \bar \gamma \dot{\bbox{R}}
+
\nabla V(\bbox{R})
\right]
\nonumber\\
& & +
i \int dt4\overline D \bbox{r}^2
\label{local_action}
\end{eqnarray}
with
\begin{eqnarray}
\bar \gamma
&=&
\lim_{\omega \rightarrow 0}
\omega {\cal I}m \Gamma(\omega)
\nonumber\\
\overline D
&=&
2 \bar \gamma T
\label{local_gamma_and_D}
\end{eqnarray}
The term in the action $i 4 \overline D \bbox{r}^2 $ 
restricts $\bbox{r}$ to have small amplitude, $\langle \bbox{r}^2 \rangle
=1/8 \overline D$. This justifies the gradient expansion of the potential 
term that was made previously in going from Eq.(\ref{Keldysh1}) to 
Eq.(\ref{Keldysh2}). The final step in the derivation of the 
Langevin equation is to integrate out the quantum/relative coordinate, 
$\bbox{r}$. The result of this simple Gaussian integration is
\begin{equation}
{\cal S}[\bbox{R}]
=
i \int_0^t dt'
\frac{ 
\left[
2 \pi \hbar \bar \rho  \hat{\bbox{z}} \times \dot{\bbox{R}}
+
2 \bar \gamma  \dot{\bbox{R}}
+
\nabla V(\bbox{R})
\right]^2}{4 \overline D}.
\label{R_action}
\end{equation}
The Skyrmion motion described by Eq.(\ref{R_action}) is equivalent to 
that described by the Langevin equation Eq.(\ref{Langevin_equation})

\end{appendix}
\widetext

\end{document}